\documentclass[conference]{IEEEtran}
\IEEEoverridecommandlockouts
\usepackage{cite}
\usepackage{amsmath,amssymb,amsfonts}
\usepackage{algorithmic}
\usepackage{graphicx}
\usepackage{textcomp}
\usepackage{xcolor}
\def\BibTeX{{\rm B\kern-.05em{\sc i\kern-.025em b}\kern-.08em
    T\kern-.1667em\lower.7ex\hbox{E}\kern-.125emX}}
\begin{document}

\title{Wireless VR/Haptic Open Platform for Multimodal Teleoperation
}

\author{
    \IEEEauthorblockN{Tae Hun Jung\IEEEauthorrefmark{1}, Hanju Yoo\IEEEauthorrefmark{1}, Yuna Jin\IEEEauthorrefmark{2}, Chae Eun Rhee\IEEEauthorrefmark{2}, and Chan-Byoung Chae\IEEEauthorrefmark{1}}
    \IEEEauthorblockA{\IEEEauthorrefmark{1}School of Integrated Technology, Yonsei University, Korea
    \\\{taehun.jung, hanju.yoo, cbchae\}@yonsei.ac.kr}
    \IEEEauthorblockA{\IEEEauthorrefmark{2}Department of Information and Communication Engineering, Inha University, Korea
    \\\ yn-j@inha.edu, chae.rhee@inha.ac.kr}
}

\maketitle

\begin{abstract}
With emerging trends in the fifth generation and robotics, the Internet of Skills will enable us to deliver skills or expertise anywhere over the Internet.
In this paper, we propose a wireless connected virtual reality and haptic communication open platform to show the proof of concept for multimodal teleoperation systems in real-time.
We focus on a practical implementation with commercial products to facilitate the access and modification of the system.
The performance of the system is measured in terms of system latency and user-centric metrics.\\

\end{abstract}

\begin{IEEEkeywords}
Internet of Skills, remote control, multimodal teleoperation, virtual reality, haptic communication.
\end{IEEEkeywords}

\section{Introduction}
The Internet of Skills (IoS) is one of the most promising use cases of the fifth generation (5G) wireless communications.
The concept of the IoS is a process of technology that can help people to access skills and expertise over the Internet regardless of the physical distance between them even on a global scale.
By deploying the properly dedicated machines in certain areas, such as smart home service, industrial manufacturing, medical assistance, and even space exploration, users are able to not only share their skills but also benefit in terms of low costs and risks \cite{dohler2017internet}.

\begin{figure}[t]
    \centering
    \includegraphics[width=0.5\textwidth]{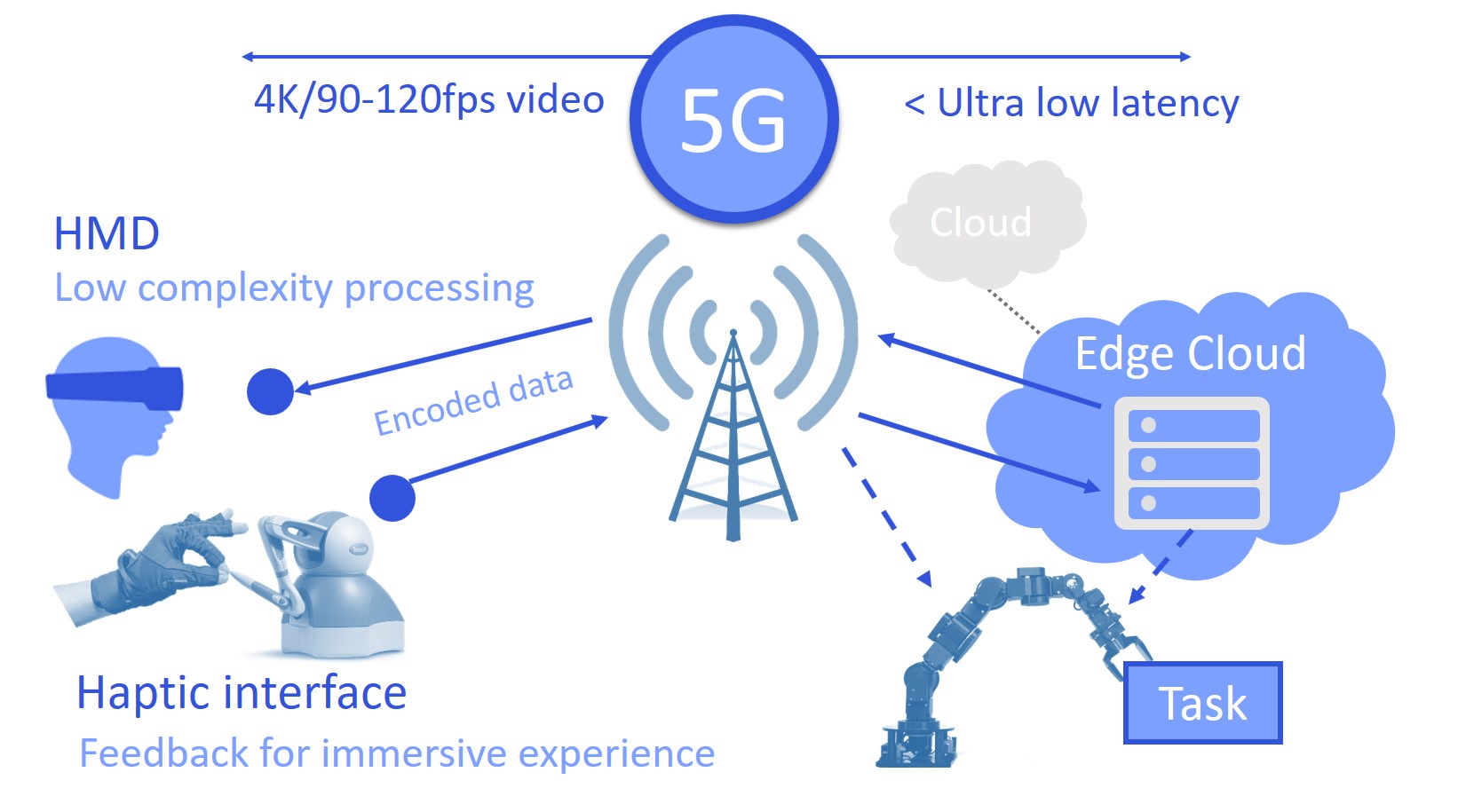}
    \caption{A description of a general scenario of the Internet of Skills}
    \label{fig:usecase}
\end{figure}

To enable these applications, a powerful platform for the remote control of machines, also known as teleoperation, should be built on the wireless communication infrastructure.
The platform needs to be capable of exchanging haptic information as well as visual and auditory data between operators and teleoperators at the same time to enable users to experience fully immersive services \cite{cizmeci2017multiplexing}.
Thanks to the development of 5G systems, especially the Tactile Internet service, several teleoperation demonstrations have been shown to the public by global telecommunication corporations \cite{lema20175g} (see Fig. \ref{fig:usecase}).
However, the current communication systems still have a few limitations such as the optimal network framework, end-to-end latency, and multimodal multiplexing, which make it difficult to fully support wireless teleoperation.

To solve these problems, we propose a "Wireless VR/Haptic Open Platform for Multimodal Teleoperation" in this paper.
The main contribution of our work is to provide a wireless teleoperation platform integrating virtual reality (VR), haptic communication and software-defined radio (SDR) implementation.
Notably, we investigate the system latency with the proposed latency reducing techniques and evaluate its performance.

\begin{figure}[t]
    \centering
    \includegraphics[width=0.5\textwidth]{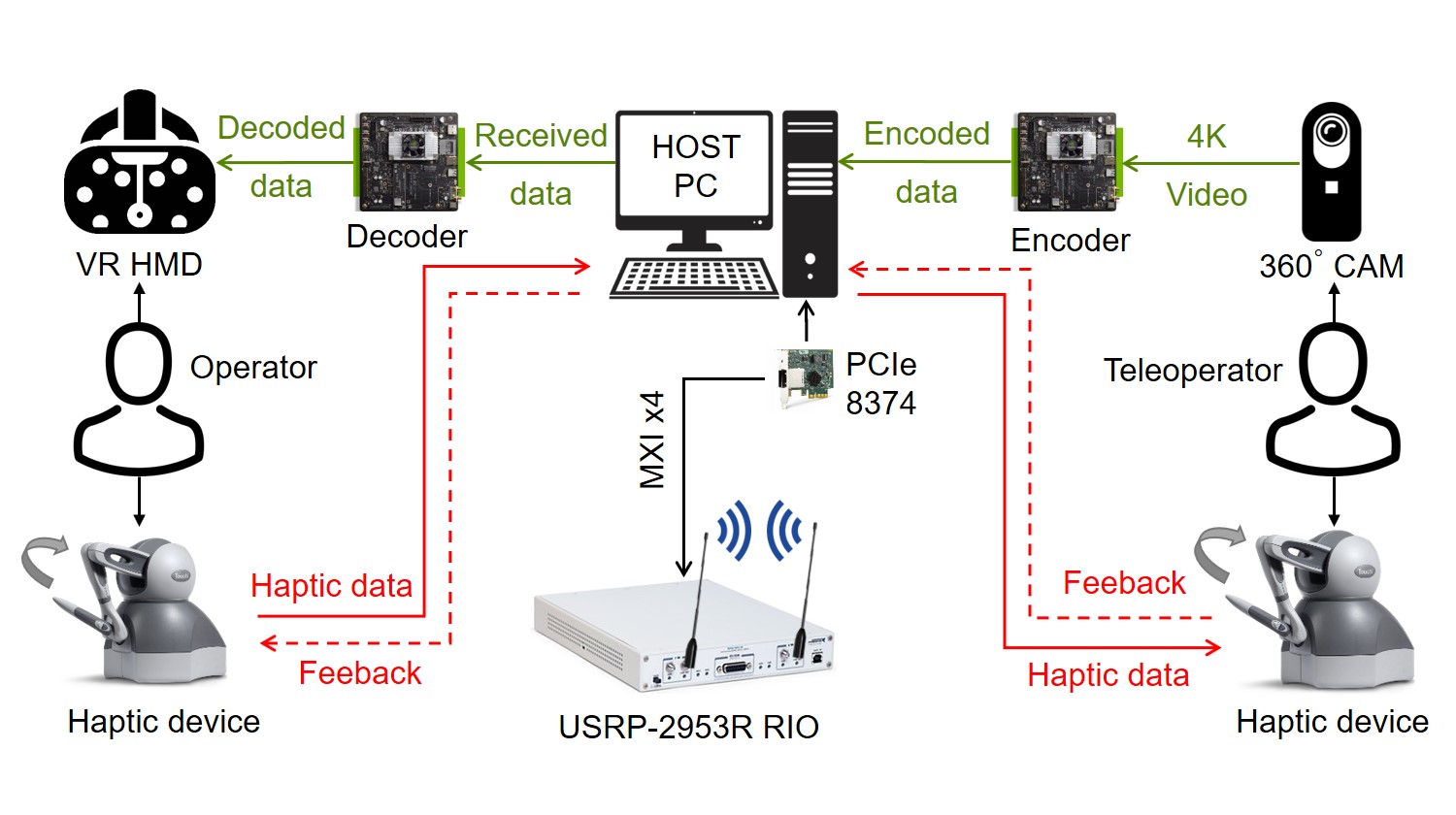}
    \caption{Structure of Wireless VR/Haptic Open Platform for Multimodal Teleoperation}
    \label{fig:demo description}
\end{figure}


\section{System Setup}

\subsection{Scenario description}

\begin{figure*}[t]
    \centering
    \includegraphics[width=1\textwidth]{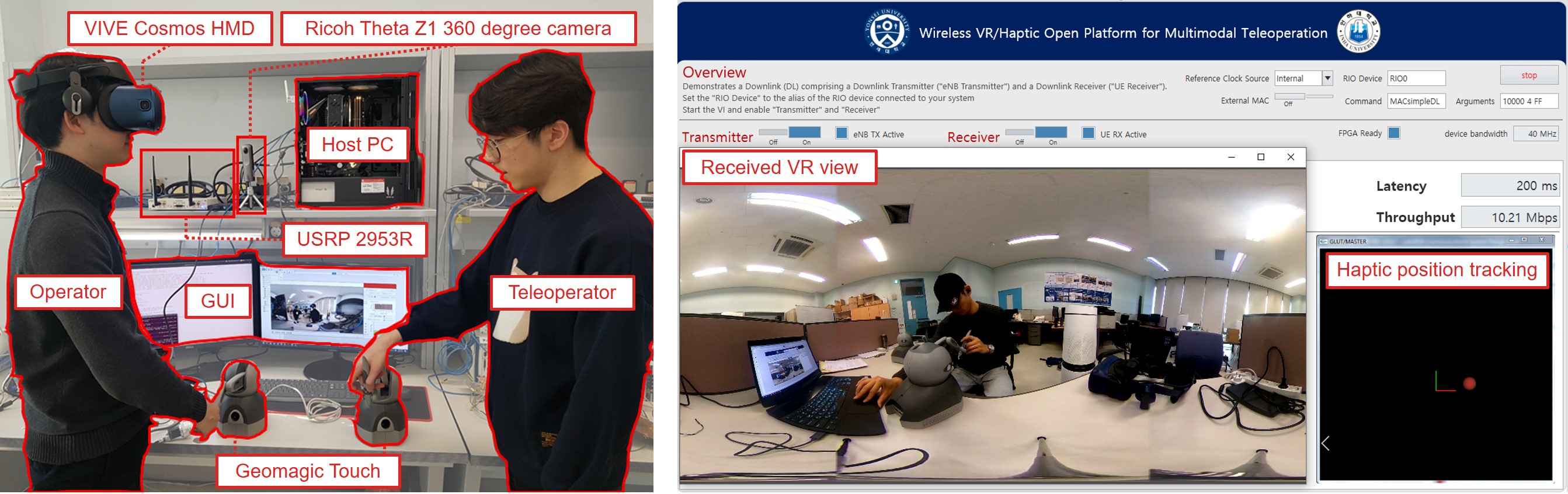}
    \caption{Experimental setup (left) and screenshot of the graphical user interface of host PC (right).}
    \label{fig:experiments}
\end{figure*}
The system structure of the proposed platform is shown in Fig. \ref{fig:demo description}.
When the operator manipulates the master robot arm, the slave robot arm copies the movements at the teleoperator side.
The camera captures and streams the view of the teleoperator's surroundings in real-time to help the operator perform the assigned task without colliding with obstacles.
If there is an unexpected situation like a collision, the operator can feel haptic feedback and adjust the movements using visual feedback.

\subsection{Hardware setup}

The operator part consists of a VR head-mounted display (HMD) and a six degree-of-freedom (DoF) robot arm, which are the interfaces for providing visual/auditory feedback and generating haptic motion data respectively.
To collect the visual/auditory information of the surroundings at the teleoperator side, the teleoperator part uses a 360-degree camera providing a full range of the vision to the operator side.
The captured video data is encoded by the embedded computing device and packetized at the host PC to be sent over the wireless link.
Moreover, the haptic motion data which is the three-dimensional position of the tips of the robot arms, is exchanged over the wireless link bi-directionally to provide the operator haptic feedback.
The wireless communication system is implemented using LabVIEW system design software and the USRP SDR platform.
The host PC performs some computational tasks and displays a graphical user interface.
More detailed information on all equipment is provided in Fig. \ref{fig:experiments}.

\subsection{System latency}
The system loop consists of several delays.
Each modality input needs to be encoded into the compressed data or, inversely, incur the encoding or decoding delay.
The encoded data is sent and received via the Internet in a User Datagram Protocol (UDP) data stream, causing a network delay.
Moreover, the real-time video and the movement of the robot should be multiplexed to be sent through the same communication link, causing a multiplexing depending on the multiplexing scheme.
On USRP, a physical delay is presented by transmitting an RF signal according to its transmission scheme.

For video data, the camera input image is divided into tiles, and the tiles are dynamically allocated to multiple encoders/decoders in consideration of computation time according to importance.
Our system assumes that the haptic motion region is a quality- and latency-sensitive region of interest (ROI) from the user's perspective.
On the encoder side, more encoder resources are allocated to the ROI tile than other tiles so that they can be encoded with sufficient time.
Therefore, various configurations are attempted to find the optimal encoding mode to have a higher image quality with the same bitrate.
The decoder side allocates more decoder resources to the ROI tile so that it can be decoded preferentially over the other tiles.
Thus, latency-sensitive areas can be updated without delay on the user's display.

\section{Conclusion}
In this paper, we presented the multimodal teleoperation system combined with wireless communications on the SDR platform.
The VR interface and the kinesthetic robot interface were used for exchanging video/audio and haptic data, respectively.
This work showed the feasibility and accessibility of our platform as well as the novel techniques to reduce the system latency, which causes severe performance degradation.
We expect our open platform to be utilized for prototyping and evaluating key technologies in 5G teleoperation scenarios.

\section*{Acknowledgment}
This research was supported by the MSIT(Ministry of Science and ICT), Korea, under the “ICT Consilience Creative Program” (IITP-2019-2017-0-01015) supervised by the IITP(Institute for Information \& communications Technology  Planning \& Evaluation) and Institute for Information \& communications Technology Planning \& Evaluation (IITP) grant funded by the Korea government(MSIT) (No.2018-0-00170, Virtual Presence in Moving Objects through 5G)

\bibliographystyle{IEEEtran}
\bibliography{IEEEabrv,2020_VRHapticDemo_arxiv}

\begin{thebibliography}{1}
\providecommand{\url}[1]{#1}
\csname url@samestyle\endcsname
\providecommand{\newblock}{\relax}
\providecommand{\bibinfo}[2]{#2}
\providecommand{\BIBentrySTDinterwordspacing}{\spaceskip=0pt\relax}
\providecommand{\BIBentryALTinterwordstretchfactor}{4}
\providecommand{\BIBentryALTinterwordspacing}{\spaceskip=\fontdimen2\font plus
\BIBentryALTinterwordstretchfactor\fontdimen3\font minus
  \fontdimen4\font\relax}
\providecommand{\BIBforeignlanguage}[2]{{%
\expandafter\ifx\csname l@#1\endcsname\relax
\typeout{** WARNING: IEEEtran.bst: No hyphenation pattern has been}%
\typeout{** loaded for the language `#1'. Using the pattern for}%
\typeout{** the default language instead.}%
\else
\language=\csname l@#1\endcsname
\fi
#2}}
\providecommand{\BIBdecl}{\relax}
\BIBdecl

\bibitem{dohler2017internet}
M.~{Dohler}, T.~{Mahmoodi}, M.~A. {Lema}, M.~{Condoluci}, F.~{Sardis},
  K.~{Antonakoglou}, and H.~{Aghvami}, ``Internet of skills, where robotics
  meets ai, 5g and the tactile internet,'' in \emph{Proc. Eur. Conf. Netw.
  Commun.}, Jun 2017, pp. 1--5.

\bibitem{cizmeci2017multiplexing}
B.~Cizmeci, X.~Xu, R.~Chaudhari, C.~Bachhuber, N.~Alt, and E.~Steinbach, ``A
  multiplexing scheme for multimodal teleoperation,'' \emph{ACM Trans.
  Multimedia Comput. Commun. Appl.}, vol.~13, no.~2, p.~21, Apr 2017.

\bibitem{lema20175g}
M.~A. {Lema}, K.~{Antonakoglou}, F.~{Sardis}, N.~{Sornkarn}, M.~{Condoluci},
  T.~{Mahmoodi}, and M.~{Dohler}, ``5g case study of internet of skills:
  Slicing the human senses,'' in \emph{Proc. Eur. Conf. Netw. Commun.}, Jun
  2017, pp. 1--6.

\end{thebibliography}

\end{document}